\documentclass{elsart}

\usepackage{amssymb}

\journal{Theoretical Computer Science}

\begin{document}

\newcommand{\sdot}{{\scriptstyle \ldots}\;}

\begin{frontmatter}

  \title{Universality and Decidability of \\
         Number-Conserving Cellular Automata }

  \author{Andr\'es Moreira}

  \ead{anmoreir@dim.uchile.cl}

  \address{Center for Mathematical Modeling and Departamento 
           de Ingenier\'{\i}a Matem\'atica \\
           FCFM, U. de Chile, Casilla 170/3-Correo 3, Santiago, Chile}

  \begin{abstract}
    Number-conserving cellular automata (NCCA) are particularly interesting, 
	 both because of their natural appearance as models of real systems, and 
	 because of the strong restrictions that number-conservation implies. Here 
	 we extend the definition of the property to include cellular automata with 
	 any set of states in $\Zset$, and show that they can be always extended to 
	 ``usual'' NCCA with contiguous states. We show a way to simulate any one 
	 dimensional CA through a one dimensional NCCA, proving the existence
	 of intrinsically universal NCCA. Finally, we give an algorithm to decide, 
	 given a CA, if its states can be labeled with integers to produce a NCCA, 
	 and to find this relabeling if the answer is positive.
  \end{abstract}

\end{frontmatter}

\section{Introduction}

  A cellular automaton (CA) is a discrete dynamical system, where the nodes
  (``cells'') of some regular lattice (usually $\Zset^d$) are mapped
  to a finite set of states; the state of a cell at a time $t+1$ is
  determined by a local function that takes at inputs the states of the cell
  and its neighbors at $t$.
  Cellular automata have been widely used as models of dynamical systems 
  in which the behavior is determined by local interaction between spatially
  fixed elements. An interesting particular class of CA is the class of 
  {\em number-conserving} CA (NCCA): roughly speaking, these are CA where the 
  states are represented as numbers, and the sum of the states over all cells
  remains constant when the states are updated.
  This conservation may help to prove some properties of the 
  dynamics, and can be usually related to the conservation of some quantity
  in the system that is being modeled. In CA traffic models (\cite{nagel}), 
  for instance, states are interpreted as the number of 
  indestructible particles located in a cell. In fact, an interpretation in 
  terms of particles can be given to any NCCA \cite{moreira1}.

  In \cite{bocc1}, Boccara and Fuk\'s give a necessary and sufficient
  condition for a one-dimensional CA of two states to be number-conserving,
  and study all the NCCA with neighborhoods $\{l,\ldots,r\}$, $l+r\leq 4$.
  In \cite{bocc2}, they give a necessary and sufficient condition that holds
  for one-dimensional CA of any number of states, and use it to study all the
  three-state NCCA for $l+r\leq 2$. In \cite{durand} Durand {\em
  et al.} formalize three different definitions of number-conservation and 
  show their equivalence. They write the generalization of Boccara's condition 
  to two dimensions, and hint on the $d$-dimensional case; the decidability of 
  number-conservation is thus proved (provided that the numeric value of
  the states is given). 
  Durand {\em et al.} also give some examples
  of NCCA in several classes of one-dimensional CA, intersecting the
  classifications of K\r{u}rka (\cite{kurka}) and Braga {\em et al.}
  (\cite{braga1,braga2}), and prove the emptiness of the remaining classes.
  In \cite{morita1} Morita and Imai prove the universality of the
  class of number-conserving reversible partitioned CA; in \cite{morita2}
  Morita {\em et al.} embed a simple general computer in a reversible,
  number-conserving two-dimensional partitioned CA. We must remark that,
  despite the title of \cite{morita1}, these articles do not settle the 
  universality of NCCA; partitioned CA can be recoded as normal CA, but 
  the recoding does not, in general, preserve number-conservation.

  In section \ref{sec:def} we give the definition of NCCA and recall
  the necessary and sufficient conditions of \cite{bocc2} and \cite{durand};
  we generalize the definitions and the conditions to allow any finite subset of
  $\Zset$ as the set of states. In section \ref{sec:ext}, we show how any
  NCCA with such set of states can be extended to a NCCA with a set
  of states of the usual form $\{0,1,\ldots,q\}$. In section \ref{sec:uni}
  we show that any one dimensional CA may be simulated by a one dimensional
  CA. This implies that NCCA are capable of universal computation, and the
  particular form of the simulation implies the even stronger property
  of intrinsical universality. Finally,
  section \ref{sec:dec} addresses the question of deciding, for any CA, whether
  its states can be relabeled with different values in $\Zset$ to make the
  CA number-conserving, thus showing the decidability of the
  number-conservation property in a wide sense.

\section{Definitions, previous results, and generalization}
\label{sec:def}

  The definition and results for NCCA have so far assumed a set of 
  states of the form $S=\{0,1,\ldots,q\}$. This makes sense for some
  applications, but in the general case, there is no reason for restricting
  the class this way: we will define them for any finite $S\subset \Zset$.
  As we will see, this doesn't change the previous results, and turns out
  to be useful in the later sections.

  \paragraph*{Cellular automata: }
  A {\em cellular automaton} $F$ in the $d$-dimensional space is
  formally described
  by a tuple $F=(d,S,N,f)$, where $d\in\Nset$ is the dimension,
  $S$ is a finite set of {\em states}, $N\subset \Zset$ is a finite 
  {\em neighborhood}, and $f:S^N\rightarrow S$ is a {\em local transition rule}%
  \footnote{
  We restrict the definitions to what we use;
  for general theory of CA, consult \cite{ca}.}.
  Since the neighborhood can always be enlarged by ignoring additional neighbors,
  we may assume it to be an hypercube, say,
  \[ 
   N=\{(a_1,\ldots,a_d)\subset \Zset^d: -l_i \leq a_i \leq r_i 
	\quad \forall \, i\}
  \]
  for some non-negative integers $l_1, r_1,\ldots, l_d, r_d$.
  A {\em configuration} is an element $c\in S^{\Zset^d}$. The dynamics of the
  system is discrete in time, and at each time step, the current configuration 
  $c^t\in S^{\Zset^d}$ determines the next one, $c^{t+1}$,
  through
  \[
   c^{t+1}_{(i_1,\ldots,i_d)} \;=\; f( c_{|(i_1,\ldots,i_d)+N}^t )
  \]
  In this way the local function induces a global one, which we will denote
  with the name of the automaton: $F:S^{\Zset^d} \rightarrow S^{\Zset^d}$.

  For the next definitions we follow Ollinger \cite{oll}; his
  formalization of intrinsic universality follows an idea presented
  in Albert and \v{C}ulik II \cite{alb} and in Bartlett and Garz\'on \cite{bart}.

  \paragraph*{Sub-automaton: }
  Let $F$ and $G$ be two cellular automata with sets of states $S_F$ and
  $S_G$ respectively; we say that $F$ is a {\em sub-automaton} of $G$ if
  there is an injective map $\phi:S_F\rightarrow S_G$ such that
  $\phi\circ F (c) = G\circ \phi (c)$ for all $c\in S_F^{\Zset^d}$. In other
  words, $F$ is just the restriction of $G$ to some configurations (up to
  state relabeling).

  \paragraph*{Scaling: }
  Here we consider the case with $d=1$. Let $\sigma$ be the {\em shift}
  function, i.e., the function $\sigma:S^\Zset\rightarrow S^\Zset$ such that
  $\sigma(c)_i=c_{i+1}$. For some strictly positive integer $m$,
  the packing map $o^m$ is the function $o^m:S^\Zset\rightarrow (S^m)^\Zset$
  such that $o^m(c)_i = (c_{mi},\ldots,c_{mi+m-1})$. Notice that both
  functions are bijective. If $F$ is a CA with states $S$, a 
  $\langle m,n,k\rangle$-rescaling of $F$ is a cellular automaton 
  $F^{\langle m,n,k\rangle}$ with states $S^m$ which verifies
  $F^{\langle m,n,k\rangle}(c)=\sigma^k\circ o^m \circ F^n \circ (o^m)^{-1} (c)$
  for all $c\in (S^m)^\Zset$.

  \paragraph*{Simulation and universality: }
  We say that a CA $F$ {\em simulates} a CA $G$ if $G$ is a sub-automaton
  of some rescaling of $F$. A CA is said to be {\em intrinsically
  universal} if it simulates any other CA. This definition, thanks to the
  restricted definition of simulation, is stronger than the usual definition 
  of computational universality, which asks for the ability to simulate a
  universal Turing machine.

%
  \paragraph*{Period: }
  A {\em period} $p \in \Nset^d$ for a configuration $c\in S^{\Zset^d}$ is a vector 
  such that $c_{i+p(c)}=c_i$, $\forall i\in \Zset^d$. It can be easily checked
  that a period is preserved trough the iterations of a CA: for any
  $F=(d,S,N,f)$, a period $p$ for $c$ will be a period for $c'=F(c)$. 
  The expression $0\leq k\leq p$ will denote the set of vectors 
  $\{0,\ldots,p_1\}\times \ldots \times \{0,\ldots,p_d\}$.

  \paragraph*{Number conservation: }
  Let $C_P(d,S)$ be the set of all the configurations in $S^{\Zset^d}$ that admit 
  a period (the --spatially-- {\em periodic configurations}); for each
  $c\in C_P(d,S)$ choose a period $p(c)$. We say that a CA $F=(d,S,N,f)$ is 
  {\em number-conserving} iff
  \begin{equation}
    \sum_{0\leq k \leq p(c)} c_k \;=\; \sum_{0\leq k \leq p(c)} F(c)_k
	 \qquad \forall c\in C_P
	\label{eq:defncca}
  \end{equation}

  Durand {\em et al.} consider CAs with $S$ of the form $\{0,\ldots,q\}$,
  for which they discuss three different definitions
  of number-conservation, and show that the three are equivalent. 
  The first is the one we just gave; the second asks that the sum over 
  all $\Zset^d$ be conserved, for all finite configurations (configurations 
  where $c_i=0$ for all but a finite number of $i\in \Zset^d$). 
  The third definition asks, for all $c\in S^{\Zset^d}$, that
  \[
   \lim_{n\rightarrow \infty} \frac{\mu_n(c)}{\mu_n(F(c))} = 1
	\qquad \, \qquad
	\textrm{ where }
	\mu_n(c)=\sum_{i\in \{-n,\ldots,n\}^d} c_i
  \]

  \subsection*{Necessary and sufficient conditions: }

  Two previous results that we will use are the necessary and
  sufficient conditions for a CA to be number-conserving in
  one dimension (proved in \cite{bocc2}) and in two dimensions
  (proved in \cite{durand}). Both assume a set of states of the form 
  $S=\{0,\ldots,q\}$, and explicitly include $0$ in the equation. 
  In fact, the only property of $0$ which is used is its {\em quiescence}: 
  we say that a state $s$ is quiescent iff $f(\{s\}^N)=s$. It 
  follows directly from the definition of number-conservation that all 
  the states of a number-conserving CA are quiescent.

  In order to show that $0\in S$ is not required in those results, we will
  deduce the result for $d=1$ again, without that condition. 
  Consider $F=(1,S,N,f)$, with $N=\{-l,\ldots,r\}$. Let $n$ be $n=l+r+1$,
  let $a\in S$ be any state, and
  take any $(x_1,\ldots,x_n) \in S^n$. Consider the configuration $c$ 
  consisting of infinite repetitions of
  \[
   x_1,x_2,\ldots,x_n,\underbrace{a,\ldots,a}_{n-1}
  \]
  If we apply equation (\ref{eq:defncca}) to this configuration (for
  period $2n-1$), we obtain
  \[
   (n-1)\;a + \sum_{k=1}^n x_k \;=\;
	 \sum_{k=1}^{n-1} f(\underbrace{a,\sdot,a}_{n-k},x_1,\sdot,x_k)
	+\sum_{k=1}^n f(x_k,\sdot,x_n,\underbrace{a,\sdot,a}_{k-1})
  \]
  Replacing $x_1=a$ in this equation, we get
  \begin{eqnarray}
   n\,a + \sum_{k=2}^n x_k & \;=\; &
	 \sum_{k=1}^{n-1} f(\underbrace{a,\sdot,a}_{n-k+1},x_2,\sdot,x_k) 
	+f(a,x_2,\sdot,x_n) \nonumber \\ & &
	+\sum_{k=2}^n f(x_k,\sdot,x_n,\underbrace{a,\sdot,a}_{k-1}) \nonumber \\
	&\;=\;&
	 \sum_{k=1}^{n-2} f(\underbrace{a,\sdot,a}_{n-k},x_2,\sdot,x_{k+1}) 
	+f(a,x_2,\sdot,x_n) \nonumber \\ & &
	+\sum_{k=2}^n f(x_k,\sdot,x_n,\underbrace{a,\sdot,a}_{k-1}) + a
	\nonumber
  \end{eqnarray}
  where we used the fact that $f(a,\ldots,a)=a$. 
  Taking the difference between the two last equations, we have
  \begin{eqnarray}
   x_1 - a & \;=\; &
	\sum_{k=1}^{n-2}
	 f(\underbrace{a,\sdot,a}_{n-k},x_1,\sdot,x_k)-
	 f(\underbrace{a,\sdot,a}_{n-k},x_2,\sdot,x_{k+1})
	\nonumber \\ & &
	+f(a,x_1,\sdot,x_{n-1}) -f(a,x_2,\sdot,x_n) + f(x_1,\sdot,x_n) - a
	\nonumber
  \end{eqnarray}
  i.e.,
   \begin{equation}
	f(x_1,\sdot,x_n) = x_1+\sum_{k=1}^{n-1}
	 f(\underbrace{a,\sdot,a}_{n-k},x_2,\sdot,x_{k+1})-
	 f(\underbrace{a,\sdot,a}_{n-k},x_1,\sdot,x_k)
	 \label{eq:ncsc3}
	\end{equation}

  This is a necessary and sufficient condition. If $F$ verifies
  equation (\ref{eq:ncsc3}) for all $(x_1,\ldots,x_n) \in S^n$, then
  it verifies (\ref{eq:defncca}): the terms in the brackets will cancel
  when the sum ranges over the whole configuration, leaving just
  the sum of states before (on the left) and after (on the right)
  the application of $F$. Hence, the theorem reads:

  \begin{thm}
  Let $F=(1,S,N,f)$ be a CA with $S\subset \Zset$, $N=\{-l,\ldots,r\}$, 
  $n=l+r+1$. Let $a$ be any state in $S$. Then, 
  $F$ is number-conserving if and only if $f$ verifies equation (\ref{eq:ncsc3})
  for all $(x_1,\ldots,x_n) \in S^n$.
  \label{teo:cns1}
  \end{thm}

  The analogous condition for two dimensions was proved in \cite{durand}
  (for $a=0$). The explicit writing of the proof is cumbersome, but the idea
  is the same as before: this time we consider a configuration with a matrix 
  $(x_{i,j})_{i=1,\ldots,m,j=1,\ldots,n}$ (the size of the neighborhood) 
  surrounded by $a's$, and write the necessary condition. Then, instead of 
  subtracting the evaluation with $x_1=a$ as before, we subtract the 
  evaluation with $x_{1,\bullet}=a$ and the evaluation with $x_{\bullet,1}=a$,
  and add the evaluation with both $x_{1,\bullet}=a$ and $x_{\bullet,1}=a$.
  It is easy to see how the procedure can be further modified to get necessary
  and sufficient conditions for $d>2$. For two dimensions, the resulting condition
  can be written as follows:

  \begin{thm}
  Let $F=(2,S,N,f)$ be a CA with $S\subset \Zset$, 
  $N=\{-l_1,\ldots,r_1\}\times\{-l_2,\ldots,r_2\}$, $n=l_1+r_1+1$,
  $m=l_2+r_2+1$. Let $a$ be any state in $S$. Then $F$ is number-conserving
  if and only if, for all $(x_{1,1},...x_{m,n}) \in S^{nm}$, it satisfies
  \begin{eqnarray} & &
	f(M_{1,1,m,n})\;= \hfill \, \nonumber \\ & &
   x_{1,1} 
	\;+ \;\sum_{i=1}^{m-1}  f(M_{2,1,i+1,n}) - f(M_{1,1,i,n})
	   +\;\sum_{j=1}^{n-1}  f(M_{1,2,m,j+1}) - f(M_{1,1,m,j}) 
   \nonumber \\ & &
	   +\;\sum_{i=1}^{m-1} \sum_{j=1}^{n-1} 
		f(M_{1,2,i,j+1}) + f(M_{2,1,i+1,j}) - f(M_{1,1,i,j}) - f(M_{2,2,i+1,j+1})
		\label{eq:midur}
   \end{eqnarray}
  where $M_{T,L,B,R}$ represents a matrix filled with $a$'s but for the
  bottom left corner, which is occupied with 
  $(x_{i,j})_{i=T,\ldots,B,j=L,\ldots,R}$.
 \end{thm}

\section{Extension from $S\subset\Zset$ to $\{0,\ldots,q\}$}
\label{sec:ext}

 The theorem of this section shows that the introduction of general $S\subset\Zset$
 as possible sets of states does not change the class of NCCA in any dramatic
 way: anything that can be seen in a NCCA with $S\in\Zset$, 
 can be seen in a NCCA with $\widetilde{S}$ of the form $\{0,\ldots,q\}$, 
 with the appropriate initial conditions. The theorem is stated and proved
 in one dimension, but we remark that versions of it for higher dimensions 
 can be proved with similar arguments.

\begin{thm}\label{teo:complete}
  Let $F$ be a NCCA with states $S\subset \Zset$ and neighborhood of
  size $n$. Then $F$ is a sub-automaton of a NCCA $\widetilde{F}$
  with states $\widetilde{S}=\{0,\ldots,\max S-\min S\}$ and neighborhood
  of size $2n$.
\end{thm}

\begin{pf*}{Proof.}
  Let $F$ be a NCCA, $F=(1,S,N,f)$, with $N=\{-l,\ldots,r\}$.
  By subtracting $\min S$ from all the states in $S$, we can assume
  that $S\subseteq\{0,\ldots,M\}$, with $M=\max S-\min S$ and $0\in S$,
  $M\in S$. In addition, without loss of generality, we may assume $l=0$:
  if $l>0$, we can apply the result to $F'=F\circ \sigma^{-l}$,
  obtaining $\widetilde{F'}$, and then take 
  $\widetilde{F}=\widetilde{F'}\circ \sigma^l$.

  We first note that for any $c\in S^{\Zset}$,
  \begin{equation}
   \sum_{j=i-r}^{i} f(c_j,\ldots,c_{j+r}) \;\leq\; rM \;+\; c_i
	\label{eq:ass}
  \end{equation}
  To prove this assertion, notice that for this to be false, $c^t_i$ must
  be strictly smaller than $M$; but in that case, we can change $c$,
  putting $c_i=M$, and we know, from the number-conservation, that the sum
  in $F(c)$ must increase accordingly. Since the only cells that can notice
  the change are those that ``see'' the state of $i$, we have that 
  $\sum_{j=i-r}^i F(c)_j$ must increase in $M-c_i$. Since the new
  sum is bounded by $(r+1)M$, we get (\ref{eq:ass}).

  The procedure to get $\tilde{F}(c)$ from a configuration $c$ is the
  following:
  \begin{enumerate}
   \item
	 $\forall i\in \Zset\, , \, c'_i=\left\{ \begin{array}{ll} c_i & \textrm{ for }c_i\in S \\ 0 & \sim \end{array}\right.$
	\item
	 $c''=F(c')$
	\item
	 $\forall i\in \Zset\, , \, c^0_i=\left\{ \begin{array}{ll} c_i'' & \textrm{ for }c_i\in S \\ c''_i+c_i & \sim \end{array}\right.$
	\item
	 For $k=1, \ldots, r$ \\
	 \hspace*{1.5em}
	  $\forall i\in \Zset\, , \, e^k_i=\max\{0,c^{k-1}_i-M\}$ \\
	 \hspace*{1.5em}
	  $\forall i\in \Zset\, , \, c^k_i=c^{k-1}_i-e^k_i+e^k_{i+1}$ 
	\item
	  $\tilde{F}(c)=c^r$
  \end{enumerate}
  So, we first remove $c_i$ from each ``invalid'' position $i$, obtaining
  $c'\in S^{\Zset}$. We apply $F$ (which is a number-conserving
  operation), and then we put $c_i$ back in its position,
  recovering the original sum of the states). At that point, 
  some $c^0_i$ may be greater than $M$. To correct this, the surplus at 
  each site is pushed to the left, and this is done $r$ times.

  We claim that $c^r$ verifies $0\leq c^r_i\leq M$, for all $i$.
  Think of the $c_i$ that are added in step 3 as particles
  labeled with $i$, and assume that each time an surplus is moved to the
  left (in step 4), preference is given to the particles with a higher
  label. Equation (\ref{eq:ass}) assures that in the $r$ sites to the left
  of $i$, there is enough place to accomodate the $c_i$ particles:
  \[
   (r+1)M-\sum_{j=i-r}^i c''_j 
	\;=\;
	M+rM-\sum_{j=i-r}^i f(c'_j,\sdot,c'_{j+r})
	\;\geq\; 
	M \;\geq\; c_i
  \]
  (remember that $c'_i=0$ for the $i$ we are considering).
  Notice that no particle is moving more than $r$ steps. Therefore,
  the final state $c^r_i$ is determined by $c''_{i+k}$ and
  $c_{i+k}$ for $k=0,\ldots,r$; i.e., it is determined by
  $c_{i+k}$ for $k=0,\ldots,2r$. To define $\tilde{f}$, we just
  consider each $(x_0,\ldots,x_{2r})\in \tilde{S}^{2r+1}$, apply the
  preceding procedure to the configuration $c$ defined by
  $c_i=x_i$ for $i=0,\ldots,2r$ and $c_i=0$ elsewhere,
  and set $\tilde{f}(x_0,\ldots,x_{2r})=c^r_0$.
  If $(x_0,\ldots,x_{2r})\in S^{2r+1}$, then $c'=c$ in the procedure,
  there are no surplusses to move, and $\tilde{f}(x_0,\ldots,x_{2r})=
  f(x_0,\ldots,x_{r})$.
\qed\end{pf*}
  
\section{Universality}
\label{sec:uni}

 \begin{thm}
  Let $F$ be a CA with $q$ states and neighborhood of size $n$. Then
  $F$ can be simulated by a NCCA $G$ with states $\{0,\ldots,2q+1\}$
  and neighborhood of size $2n$.
 \end{thm}

 \begin{pf*}{Proof.}
  Let $F$ be $F=(1,S,N,f)$, with $S=\{1,\ldots,q\}$ and $N=\{-l,\ldots,r\}$.
  We define $\hat{F}=(1,\hat{S},\hat{N},\hat{f})$ with
  \[
   \hat{S}=\{-q,\ldots,-1,0,1,\ldots,q\}
	\qquad , \qquad 
	\hat{N}=\{-2l-1,\ldots,2r+1\}
  \]
  and $\hat{f}(a_{-2l-1},\ldots,a_{2r+1})$ given by
  {\small
   \[
     \left\{
	   \begin{array}{ll}
	    f(a_{-2l},a_{-2l+2},\sdot,a_{2r}) &
	     \textrm{ if } a_{2i} = -a_{2i+1} > 0 \, , \, -l\leq i\leq r \\
	    f(-a_{-2l},-a_{-2l+2},\sdot,-a_{2r}) &
	     \textrm{ if } a_{2i} = -a_{2i-1} < 0 \, , \, -l\leq i\leq r \\
		 a_0 &
		  \textrm{ otherwise }
	  \end{array}
	 \right.
  \]}
  The idea is the following: each cell is split in two, and
  the new cells are occupied with a negative and a positive copy of the
  state the cell had. The iterations will follow the original CA rule, both
  on the negative and on the positive cells, and the sum will remain
  constant (zero). 
  
  With the definition given above, a positive cell will change its state only if it sees a ``correct 
  configuration'' around it (a configuration of the form $a,-a,b,-b,\ldots$),
  and sees that its right neighbor is seeing it too; similarly, a negative
  cell will change only if it sees that its left --positive-- neighbor will
  change in the same way (with the opposite sign). Hence, the only changes
  in the configuration are done in pairs of cells, and on each of this
  pairs the sum is conserved (and is 0). The CA $\hat{F}$ is then
  number-conserving. Through the injection $a\rightarrow (a,-a)$ $F$ becomes
  a sub-automaton of $\hat{F}^{\langle 2,1,0 \rangle}$. To avoid
  ``negative'' states, we add $q$ to all the states of $\hat{F}$.
  \qed\end{pf*}

 \begin{cor}
  There are in\-trin\-si\-cally uni\-ver\-sal one-di\-men\-sion\-al NCCA.
 \end{cor}

 \begin{pf*}{Proof.}
  The relation of simulation is a preorder \cite{oll}, and there exist
  intrinsically universal one-dimensional CA.
 \qed\end{pf*}

\section{Decidability}
\label{sec:dec}

  Equations (\ref{eq:ncsc3}) and (\ref{eq:midur}) show necessary and sufficient 
  conditions for a CA in one and two dimensions, respectively, to be 
  number-conserving; equations similar to them for higher dimensions are hard
  to write, but not to obtain. The decidabily of the property of 
  number-conservation is thus proved, but with one important restriction: 
  the numeric values of the states are taken as given. This is not a complete 
  answer, since the labeling of the states in a CA is, in principle, arbitrary. 
  If a model produces a CA with a set of states 
  defined as colors, letters, or even numbers, {\em can we relabel the states 
  with numbers such that the CA is number-conserving?}
  If the answer is yes, it would be interesting, since the conserved 
  quantity could be traced back to the original system, or could help 
  to prove some 
  property of the CA. In particular, in \cite{moreira1} it is shown that NCCA 
  can always be interpreted in terms of the interactions of indestructible 
  particles; it would be interesting, for a system with states, say, 
  {\em blue}, {\em yellow} and {\em red}, to know if the dynamics can be 
  expressed in such terms. 
%

  If the set of states is required to be of the form $\{0,\ldots,q-1\}$, 
  where $q=|S|$, then the answer to the problem 
  is easy: just consider the $q!$ possible permutations of the labels, 
  and check, for each of them, if the CA is number-conserving (using
  equations (\ref{eq:ncsc3}) or (\ref{eq:midur})). But this requirement is 
  arbitrary: perhaps what we are seeing {\em can} be interpreted in terms of 
  particles, but there is some quantity of particles which happens to never 
  occur in a cell, and is therefore {\em absent} from our current set of 
  states. In general, we would like to know if we can relabel the states of 
  the CA with $|S|$ different elements of $\Zset$, so as to make the CA 
  number-conserving.

  {\bf Example 1 }
  Consider $F=(1,S,N,f)$ with $S=\{a,b,c\}$,
  $l=r=3$, and $f$ defined by:
  {\small
  \[
   f(x_0,x_1,x_2,x_3,x_4,x_5,x_6)
	\;=\;
	\left\{ \begin{array}{llll}
	 c  & \textrm{ if } & (x_0=x_1=x_3=x_4=a \land x_2=b) \\
	    &  \lor         & (x_1=x_2=x_4=x_5=a \land x_3=b) \\
	    &  \lor         & (x_2=x_3=x_5=x_6=a \land x_4=b) \\
	x_3 & \sim          & 
	\end{array} \right.
  \]}
  It can be checked, using equation (\ref{eq:ncsc3}), that all 
  the bijections $\phi:\{a,b,c\}\rightarrow\{0,1,2\}$ produce rules that 
  are not number-conserving.
  But $F$ does become number-conserving if we relabel its states with
  \[
   \phi(a)=0 \qquad \phi(b)=3 \qquad \phi(c)= 1
  \]
  It may be interpreted in terms of particles: if three particles are in a 
  cell and the two neighboring cells in each direction are empty, then they 
  separate; one goes to the left, one to the right, and one stays in 
  the cell.

  \begin{thm}
  Let $F=(d,S,N,f)$ be a CA. Then there is an algorithm to 
  decide if there exists a relabeling 
  that makes $F$ number-conserving, and to find it if the answer is positive.
   \label{teo:dec}
  \end{thm}

  \begin{pf*}{Proof.}
  Write $S$ as $S=\{s_0,s_1,\ldots,s_{|S|-1}\}$. 
  We are looking for an injective function $\phi:S\rightarrow\Zset$, such
  that $F$, redefined with states $\phi(S)$, becomes number-conserving.
  To find it solution, we will use the equation that gives the necessary 
  and sufficient condition for the CA to be number-conserving; it will be 
  equation (\ref{eq:ncsc3}), (\ref{eq:midur}), or a form for higher dimensions, 
  depending on the dimension in which $F$ is defined. 
  In \cite{bocc2}, the equation was used to list the possible NCCAs,
  taking the values of the function $f$ as the unknown variables.
  This time, the unknown variables of the equation are not the evaluations of 
  the rule, but the numeric value of the states: the necessary and sufficient 
  condition to make the relabeled version of $f$ number-conserving is, in 
  the one-dimensional case (the other cases are analogous), that
  \begin{eqnarray}
   \phi(f(x_1,\sdot,x_n)) \;=\; \phi(x_1) + \sum_{k=1}^{n-1} &
      \phi(f(\underbrace{s_0,\sdot,s_0}_k,x_2,\sdot,x_{n-k+1}))
		\nonumber \\
	& - \phi(f(\underbrace{s_0,\sdot,s_0}_k,x_1,\sdot,x_{n-k}))
  \end{eqnarray}
  for all $(x_1,x_2,\ldots,x_n) \in S^n$, where $n=|N|$. (Notice that from
  theorem \ref{teo:cns1}, the choice of $s_0$ is arbitrary.)
  This is an homogeneous linear system of $|S|^n$ equations and $|S|$ 
  variables. The set of solutions will be a linear subspace 
  $V\in \Rset^{|S|}$. If $V=\{0\}$, the algorithm gives a negative answer.
  If $V\neq \{0\}$, we must still find out if it contains solutions 
  {\em which have all the coordinates different} (if not, then there is no 
  injection). In other words, we must check if 
  $V\setminus \left( \bigcup_{j\neq k} E_{jk} \right) \neq \emptyset$, 
  where $E_{jk}$ is the hyperplane $x_j=x_k$. But 
  \[
    V\setminus \big( \bigcup_{j\neq k} E_{jk} \big) = \emptyset
	 \iff
	 V = V\cap \big( \bigcup_{j\neq k} E_{jk} \big) = \bigcup_{j\neq k} V\cap E_{jk}
	\]\[
	 \iff
	 \exists j,k : V = V\cap E_{jk}
	 \iff
	 \exists j,k : V \subseteq E_{jk}
  \]
  where we use the fact that each $V\cap E_{jk}$ is a subspace of $V$, and a 
  linear space cannot be a finite union of proper subspaces \cite{algebra}. 
  The last condition can be easily checked by the algorithm, by adding the 
  equation $\phi(x_j)=\phi(x_k)$ to the equation system and seeing if the 
  space of solutions is the same.

  If $V\setminus \left( \bigcup_{j\neq k} E_{jk} \right) \neq \emptyset$, 
  then there is a solution $\phi$ to the problem, and it can be found in finite
  time. We can now relabel $A$ with $\phi$, making it number-conserving;
  if a set of states of the form $\{0,\ldots,q\}$ is desired, 
  theorem \ref{teo:complete} may be applied.
  \qed\end{pf*}

 Note that the algorithm in theorem \ref{teo:dec} is only needed 
 when $|S|>2$. For $|S|=2$, if
 the CA can be made number-conserving, then it will be number-conserving
 for the two possible choices of $\phi:S\rightarrow \{0,1\}$:
 if $S=\{a,b\}\subset \Zset$ is a solution, then $\frac{S-a}{b-a}=\{0,1\}$
 is one too.

 \section{Conclusions}

 In this paper we develop several notions and results related to the class
 of number-conserving cellular automata, generalizing the definition
 and showing both the computational universality of the class and the
 decidability of the property that defines it (even for CA that are
 not initially expressed with numerical states).

 The generalization of the definition allows the CA to have any set
 of states in $\Zset$; we show that the necessary and sufficient conditions
 previously given for number-conservation do not require the previously
 assumed set of states of the form $\{0,1,\ldots,q\}$. Furthermore, the
 generalization does not change the class in any profound way: in section
 \ref{sec:ext} we show that for a NCCA with set of states $S\subset \Zset$,
 the set of states can be completed to a contigous set of states,
 extending the rule while keeping the number-conservation property. Hence,
 any ``generalized'' NCCA can be seen as a ``usual'' NCCA, restricted to
 a subset of its configurations. The extension we show in theorem 
 \ref{teo:complete} requires an enlargement of the neighborhood; since we
 do not know examples where this enlargement is really needed, it may be
 the case that the result can be improved to an extension that
 preserves the neighborhood of the CA.

 Section \ref{sec:uni} shows that any one-dimensional CA can be
 simulated by a one-dimensional NCCA; the construction is very simple
 and proves that members of the latter of the latter class exhibit
 intrinsical universality, a property that implies computational
 universality.
 
 Section \ref{sec:dec} dealed with the question of deciding, for a given
 CA, whether its states can be labeled with integers, making 
 it number-conserving. This question turns out to be decidable, and we show
 how to find the relabeling, if there is one. The particular interest
 of this result is that it may help to reveal some conservative dynamics
 in an otherwise not-conserving CA; such a conservation may
 afterwards be interpreted in terms of indestructible particles~\cite{moreira1}.

\end{document}